\documentclass[aoas,MSNbibl,nameyear,rotating,seceqn,dvips]{arximspdf}
\usepackage{mathbh,dcolumn}
\usepackage{graphicx}
\usepackage{url,breakurl}

\doi{10.1214/14-AOAS747} 
\volume{8}
\issue{4}
\pubyear{2014}
\firstpage{2150}
\lastpage{2174}
\docsubty{FLA}

\makeatletter
\newcolumntype{d}[1]{D{.}{.}{#1}}
\def\mathbbm{\mathbh}
\newtheorem{thmm}{Theorem}

\makeatother

\begin{document}
\begin{frontmatter}

\title{Imputation of truncated \textit{p}-values for meta-analysis methods and
its genomic application\thanksref{T2}}
\runtitle{Imputation in microarray meta-analysis}

\begin{aug}
\author[A]{\fnms{Shaowu}~\snm{Tang}\ead[label=e1]{sht41@pitt.edu}\thanksref{m1}},
\author[B]{\fnms{Ying}~\snm{Ding}\ead[label=e2]{yid5@pitt.edu}\thanksref{m2}},
\author[C]{\fnms{Etienne}~\snm{Sibille}\ead[label=e3]{Etienne.Sibille@camh.ca}\thanksref{m3}},
\author[D]{\fnms{Jeffrey~S.}~\snm{Mogil}\ead[label=e4]{jeffrey.mogil@mcgill.ca}\thanksref{m4}},
\author[E]{\fnms{William~R.}~\snm{Lariviere}\ead[label=e5]{lariwr@upmc.edu}\thanksref{m5}}
\and
\author[F]{\fnms{George~C.}~\snm{Tseng}\corref{}\ead[label=e6]{ctseng@pitt.edu}\thanksref{m5}}
\runauthor{S. Tang et al.}

\thankstext{T2}{Supported in part by NIH R21MH094862.}

\affiliation{Roche Molecular Systems, Inc.\thanksmark{m1},
Pfizer\thanksmark{m2},
University of Torontot\thanksmark{m3},
McGill University\thanksmark{m4} and
University of Pittsburgh\thanksmark{m5}}
\address[A]{S. Tang\\
Roche Molecular Systems, Inc.\\
4300 Hacienda Drive\\
Pleasanton, California 94588\\
USA\\
\printead{e1}}
\address[B]{Y. Ding\\
Pfizer\\
10646 Science Center Dr.\\
San Diego, California 92121\hspace*{14pt}\\
USA\\
\printead{e2}}
\address[C]{E. Sibille\\
Centre for Addiction and Mental Health\\
University of Toronto\\
250 College Street\\
Toronto, ON\\
Canada\\
\printead{e3}}
\address[D]{J.~S. Mogil\\
Department of Psychology\\
McGill University\\
1205 Dr. Penfield Avenue\\
Montreal, QC\\
Canada\\
\printead{e4}\hspace*{15pt}}
\address[E]{W.~R. Lariviere\\
Pittsburgh Center for Pain Research\\
University of Pittsburgh\\
200 Lothrop Street\\
Pittsburgh, Pennsylvania 15213\\
USA\\
\printead{e5}}
\address[F]{G. C. Tseng\\
Department of Biostatistics\\
University of Pittsburgh\\
130 Desoto Street\\
Pittsburgh, Pennsylvania 15261\\
USA\\
\printead{e6}}


\end{aug}

\received{\smonth{11} \syear{2013}}
\revised{\smonth{4} \syear{2014}}

%
\begin{abstract}
Microarray analysis to monitor expression activities in thousands of
genes simultaneously has become routine in biomedical research during
the past decade. A tremendous amount of expression profiles are
generated and stored in the public domain and information integration
by meta-analysis to detect differentially expressed (DE) genes has
become popular to obtain increased statistical power and validated
findings. Methods that aggregate transformed $p$-value evidence have been
widely used in genomic settings, among which Fisher's and Stouffer's
methods are the most popular ones. In practice, raw data and $p$-values
of DE evidence are often not available in genomic studies that are to
be combined. Instead, only the detected DE gene lists under a certain
$p$-value threshold (e.g., DE genes with $p$-value${}<0.001$) are reported in
journal publications. The truncated $p$-value information makes the
aforementioned meta-analysis methods inapplicable and researchers are
forced to apply a less efficient vote counting method or na\"{i}vely
drop the studies with incomplete information. The purpose of this paper
is to develop effective meta-analysis methods for such situations with
partially censored $p$-values. We developed and compared three imputation
methods---mean imputation, single random imputation and multiple
imputation---for a general class of evidence aggregation methods of
which Fisher's and Stouffer's methods are special examples. The null
distribution of each method was analytically derived and subsequent
inference and genomic analysis frameworks were established. Simulations
were performed to investigate the type I error, power and the control
of false discovery rate (FDR) for (correlated) gene expression data.
The proposed methods were applied to several genomic applications in
colorectal cancer, pain and liquid association analysis of major
depressive disorder (MDD). The results showed that imputation methods
outperformed existing na\"{i}ve approaches. Mean imputation and
multiple imputation methods performed the best and are recommended for
future applications.
\end{abstract}

%
\begin{keyword}
\kwd{Microarray analysis}
\kwd{meta-analysis}
\kwd{Fisher's method}
\kwd{Stouffer's method}
\kwd{missing value imputation}
\end{keyword}
\end{frontmatter}

\section{Introduction and motivation}\label{sec1}
Microarray analysis to monitor expression activities in thousands of
genes simultaneously has become routine in biomedical research during
the past decade. The rapid development in biological high-throughput
technology results in a tremendous amount of experimental data and many
data sets are available from public domains such as Gene Expression
Omnibus (GEO) and ArrayExpress. Since most microarray studies have
relatively small sample sizes and limited statistical power,
integrating information from multiple transcriptomic studies using
meta-analysis techniques is becoming popular. Microarray meta-analysis
usually refers to combining multiple transcriptomic studies for
detecting differentially expressed (DE) genes (or candidate markers).
DE gene analysis identifies genes differentially expressed across two
or more conditions (e.g., cases and controls) with statistical
significance and/or biological significance (e.g., fold change).
Microarray meta-analysis in many situations refers to performing
traditional meta-analysis techniques on each gene repeatedly and then
controlling the false discovery rate (FDR) to adjust $p$-values for
multiple comparison [\citet{Borovecki}; \citet{Cardoso};
Pirooznia, Nagarajan and Deng (\citeyear{Pirooznia}); \citet{Segal}].
Fisher's method [Fisher
(\citeyear{Fisher})] was the first meta-analysis technique introduced in microarray
data analysis in $2002$ [\citet{Rhodes}], followed by Tippett's
minimum $p$-value method in $2003$ [\citet{Moreau}]. Subsequently,
many meta-analysis approaches have been used in this field, including
extensions of existing meta-analysis techniques and novel methods to
encompass the challenges presented in the genomic setting
[\citet{Choi}, Choi et al. (\citeyear{Choi2}),
Moerau et al. (\citeyear{Moreau}), \citet{Owen}, \citet{Li}, and see a review paper by
Tseng, Ghosh and Feingold (\citeyear{Tseng2})].

To combine findings from multiple research studies, one needs
to know either the effect size or the $p$-value for each study. Since the
differences in data structures and statistical hypotheses across
multiple studies may make the direct combination of effect sizes
impossible or the result suspicious, combining $p$-values from multiple
studies is often more appealing. Popular $p$-value combination methods
[see review and comparative papers Tseng, Ghosh and Feingold (\citeyear{Tseng2})
and \citet{Chang}] can be split into two major categories of evidence aggregation
methods (including Fisher's, Stouffer's and logit methods) and order
statistic methods [such as minimum $p$-value, maximum $p$-value and $r$th
ordered $p$-value by Song and Tseng (\citeyear{Song})]. Evidence aggregation methods
utilize summation of certain transformations of $p$-values as their test
statistics to aggregate differential expression evidence across
studies. Among evidence aggregation methods, Fisher's method is the
most well known, in which the test statistic is defined as
$T^{\mathrm{Fisher}}=-2\sum_{k=1}^K \log(p_k)$, where $K$ is the number of
independent studies that are to be combined and $p_k$ is the $p$-value of
individual study $k, 1\leq k\leq K$. Under the null hypothesis of no
effect size in all studies and assuming that studies are independent
and models for assessing $p$-values are correctly specified, $T^{\mathrm{Fisher}}$
follows a chi-square distribution with degrees of freedom $2K$.
Fisher's method has been popular due to its simplicity and some
theoretical properties, including admissibility under Gaussian
assumption [Birnbaum (\citeyear{Birnbaum1,Birnbaum2})] and asymptotically Bahadur
optimality (ABO) under equal nonzero effect sizes across studies
[Littel and Folk (\citeyear{Littell1,Littell2})]. Some variations of Fisher's methods
were proposed by using unequal weights or a trimmed version of Fisher's
test statistic [\citet{Olkin}]. Another widely used evidence
aggregation method is the Stouffer's method [\citet{Stouffer}], in which
the test statistic is defined as $T^{\mathrm{Stouffer}}=\sum_{k=1}^K \Phi
^{-1}(p_k)$,\vspace*{1pt} where $\Phi^{-1}(\cdot)$ is the inverse cumulative
distribution function (CDF) of standard normal distribution.

 In order to combine $p$-values, all $p$-values across studies
should be known. In genomic applications, however, raw data and thus
$p$-values are often not available and usually only a list of
statistically significant DE genes ($p$-value less than a threshold) is
provided in the publication [Griffith, Jones and Wiseman (\citeyear{Griffith})]. Although many
journals and funding agencies have encouraged or enforced data sharing
policies, the situation has only improved moderately. Many researchers
are still concerned about data ownership, and researchers whose studies
are sponsored by private funding are not obligated to share data in the
public domain. For example, in Chan et al. (\citeyear{Chan}), publications of $23$
colorectal cancer versus normal gene expression profiling studies were
collected to perform meta-analysis to identify consistently reported
candidate disease-associated genes. However, only one raw data set is
available from the Gene Expression Omnibus (GEO;
\url{http://www.ncbi.nlm.nih.gov/geo/}, GSE3294) and most other papers only
provided a list of DE genes (and their $p$-values) under a prespecified
$p$-value threshold. A~second motivating example comes from a microarray
meta-analysis study for pain research [LaCroix-Fralish et al. (\citeyear{LaCroix})], in
which $19$ microarray studies of pain models were collected to detect
the gene signature and patterns of pain conditions. Among the $19$
studies, only one raw data set was available on the author's website
and all the other papers reported the DE gene lists under different
thresholds.

 In these two motivating examples (details to be shown in
Sections \ref{sec4.1} and \ref{sec4.2}), the incomplete data forced researchers to either
drop studies with incomplete $p$-values or apply the convenient vote
counting method [\citet{Hedges4}]. Dropping studies with
incomplete information greatly reduces the statistical power and is
obviously not applicable in the two motivating examples since the
complete data was available in only one study. The conventional vote
counting procedure is well known as flawed and low-powered [\citet{McCarley}].
Ioannidis et al. (\citeyear{Ioannidis}) attempted to reproduce $18$
microarray studies published in Nature Genetics during 2005--2006.
Interestingly, only two were ``in principle'' replicated, six ``partially''
replicated and ten could not be reproduced. This result illustrates
well the widespread difficulty of obtaining raw data or reproducing
published results in the field. Therefore, developing methods to
efficiently combine studies with truncated $p$-value information is an
important problem in microarray meta-analysis.

 In this paper, we assume that $K=K_1+K_2$ studies are
combined. In $K_1$ studies, the raw gene expression data matrix and
sample annotations are available and the complete $p$-values $p_{gi}$
($1\leq g\leq G$ for genes and $1\leq i\leq K_1$) can be reproduced for
meta-analysis. For the remaining $K_2$ studies, either the raw data or
annotation is not available. Only incomplete information of a DE gene
list (under $p$-value threshold $\alpha_i$ for study $i$) is provided in
the journal publication. In this situation, the available information
is an indicator function $\mathbbm{1}_{\{p_{gi}<\alpha_i\}}$ to
represent whether the $p$-value of gene $g$ in study $i$ is smaller than
$\alpha_i$ or not. We outline the paper structure as the following. In
Section~\ref{sec2} a general class of evidence aggregation meta-analysis
methods under a univariate scenario was investigated for the mean
imputation, the single random imputation and the multiple imputation
methods, respectively, in which the exact or approximate null
distributions were derived under the null hypotheses and the results
are shown for the Fisher and the Stouffer methods. In Section~\ref{sec3.1}
simulations of the expression profile were performed to compare
performance of different methods. Simulations were further performed in
Section~\ref{sec3.2} using 8 major depressive disorder (MDD) and 7 prostate
cancer studies where raw data were completely available and the true
best performance (complete case) could be obtained. In Section~\ref{sec4} the
proposed methods were applied to the two motivating examples. In
Section~\ref{sec4.1} the methods were applied to $7$ colorectal cancer
studies, where the raw data were available only in $3$ studies. In
Section~\ref{sec4.2} the proposed methods were applied to $11$ microarray
studies of pain conditions, where no raw data were available. In
Section~\ref{sec4.3} we developed an unconventional application of the proposed
methods to facilitate the large computational and data storage needs in
a liquid association meta-analysis. Discussions and conclusions are
included in Section~\ref{sec5} and all proofs are left in the \hyperref[app]{Appendix}.

\section{Methods and inferences}\label{sec2}

\subsection{Evidence aggregation meta-analysis methods}\label{sec2.1}
Here we consider a general class of univariate evidence aggregation
meta-analysis methods (for gene $g$ fixed), in which the test
statistics are defined as the sum of selected transformations of
$p$-values for each individual study. Without loss of generality,
assuming that $F_X(\cdot)$ is the cumulative distribution function
(CDF) of a continuous random variable $X$, the test statistic $T$ is
defined as
%
\begin{equation}\label{eq2.1}
T = \sum_{i=1}^K T_k:= \sum
_{k=1}^K F^{-1}_X(p_k),
\end{equation}
where $p_k$ is the $p$-value from the $k$th study. Theoretically, $X$ can
be any continuous random variable. However, in practice, $X$ is usually
selected such that the test statistic $T$ follows a simple
distribution. For instance, when $X\sim\chi^2_2$, it holds $T\sim\chi
^2_{2K}$ (Fisher's method) and $T\sim\mathrm{N}(0,K)$ holds, provided
$X\sim\mathrm{N}(0,1)$ (Souffer's method).

 The hypothesis that corresponds to testing the homogeneous
effect sizes of $K$ studies by evidence aggregation methods is a
union-intersection test (UIT) [\citet{Roy}]:
%
\begin{equation}\label{eq2.2}
H_0\dvtx \bigcap_{k=1}^K\{
\theta_k=0\}\quad \mbox{versus}\quad H_A\dvtx \bigcup
_{k=1}^K\{\theta_k\neq0\}.
\end{equation}

 In this paper, we focus on two popular special cases:
\begin{longlist}[1.]
\item[1.] Fisher's method [Fisher (\citeyear{Fisher})]: When $X\sim\chi^2_2$,
$T_k=F^{-1}_X(p_k)=\break  -2\log(p_k)$.
\item[2.] Stouffer's method [\citet{Stouffer}]: When $X\sim\mathrm
{N}(0,1)$, $T_k=F^{-1}_X(p_k)=\Phi^{-1}(p_k)$.
\end{longlist}

 Another example is the logit method [\citet{Hedges}], where $T_k=-\log(\frac{p_k}{1-p_k})$. But since this method is
rarely used in practice, we will not examine it further in this paper.
To apply the evidence aggregation meta-analysis methods mentioned
above, all the $p$-values should be observed. However, in genomic
applications, it often happens that $p$-values of some studies are
truncated and only their ranges are reported. Two na\"{\i}ve methods
are commonly used to overcome this situation: the vote counting method
or the available-case method which only combines studies with observed
$p$-values. The available-case method discards rich information contained
in the studies with truncated $p$-values and, therefore, the statistical
power is reduced. \citet{Hedges4} showed that the power of vote
counting converges to $0$ when many studies of moderate effect sizes
are combined and, therefore, the vote counting method should be avoided
whenever possible. In this section, three imputation methods---mean
imputation, single random imputation and multiple imputation method---are proposed and investigated to combine studies with truncated
$p$-values and the corresponding null distributions are derived
analytically, respectively. We first define some notation.

 Assume that $K$ independent studies are to be combined and
$p_1,\ldots,p_K$ are the corresponding $p$-values. Without loss of
generality, assume that all the $p$-values are available in the first
$K_1$ studies and only the indicator function of DE evidence is
reported in the other $K_2$ studies.

 Define a pair $(c_i,x_i), i=1,\ldots,K$ for each study, in
which $c_i$ is the ``censoring'' indicator satisfying
%
\begin{equation}\label{eq2.3}
c_i := \cases{ %
0, & \quad$\mbox{if }
p_i \mbox{ is observed }(\mbox{i.e.}, 1\leq i\leq K_1),$
\vspace*{2pt}\cr
1, &\quad $\mbox{if } p_i \mbox{ is censored } (\mbox{i.e.},
K_1+1\leq i\leq K),$}
\end{equation}
and $x_i$ is the final observed values which is defined as
%
\begin{equation}\label{eq2.4}
x_i := \cases{ %
 p_i, &\quad
$\mbox{if } c_i=0,$
\vspace*{2pt}\cr
\mathbbm{1}_{\{p_i<\alpha_i\}}, &\quad $\mbox{if } c_i=1,$}
\end{equation}
where $\alpha_i$ is the $p$-value threshold for study $i\ (K_1+1\leq
i\leq K_1+K_2=K)$. For each $i=1,2,\ldots,K$, one can impute the
missing value by $\tilde{p}_i$:
\[
\tilde{p}_i=p_i\cdot\mathbbm{1}_{\{c_i=0\}}+[q_i
\cdot\mathbbm{1}_{\{
x_i=1\}}+r_i\cdot\mathbbm{1}_{\{x_i=0\}}]
\cdot\mathbbm{1}_{\{c_i=1\}
}
\]
with $q_i \in(0,\alpha_i)$, and $r_i \in[\alpha_i,1)$. Sections~\ref{sec2.2}--\ref{sec2.4} develop three imputation methods for selection of $q_i$ and $r_i$.
\subsection{Mean imputation method}\label{sec2.2}
The simplest imputation method is the mean imputation method, in which
$q_i=\frac{\alpha_i}{2}$ and $r_i=\frac{1+\alpha_i}{2}$. Then the test
statistic $\tilde{T}$ for truncated data satisfies
%
\begin{eqnarray}\label{eq2.5}
\tilde{T}&=&\sum_{i=1}^K
\tilde{T}_i = \sum_{i=1}^K
F^{-1}_X(\tilde {p}_i)=\sum
_{i=1}^{K_1} F^{-1}_X(p_i)+
\sum_{j=1}^{K_2}F^{-1}_X(
\tilde {p}_{K_1+j})
\nonumber
\\[-8pt]
\\[-8pt]
\nonumber
&=&A+\sum_{j=1}^{K_2}
B_j,
\end{eqnarray}
with
%
\begin{eqnarray}
A&=&\sum_{i=1}^{K_1} F^{-1}_X(p_i)\quad
\mbox{and}
\nonumber
\\
\qquad B_j&=&F^{-1}_X(\tilde
{p}_{K_1+j})
\nonumber
\\[-8pt]
\\[-8pt]
\nonumber
&=&F^{-1}\biggl(\frac{\alpha_{K_1+j}}{2}\biggr)\cdot
\mathbbm{1}_{\{
p_{K_1+j}<\alpha_{K_1+j}\}}\\
&&{}+F^{-1}\biggl(\frac{1+\alpha_{K_1+j}}{2}\biggr)\cdot
\mathbbm{1}_{\{p_{K_1+j}\geq\alpha_{K_1+j}\}}\nonumber
\end{eqnarray}
for $j=1,\ldots,K_2$. Recall that under the null hypothesis, the random
variable $A$ satisfies $A\sim\chi^2_{2K_1}$ for the Fisher method and
$A\sim\mathrm{N}(0,K_1)$ for the Stouffer method. Obviously, $B_j$
follows a Bernoulli distribution.

 The results can be summarized into the following theorem
(proof left to Appendix \ref{appB1}):

\begin{thmm}\label{th1}
For $j=1,2,\ldots,K_2$ and given $t$, by defining
%
\begin{eqnarray}
b_j&=& F_X^{-1}\biggl(\frac{\alpha_{K_1+j}}{2}
\biggr)-F_X^{-1}\biggl(\frac{1+\alpha
_{K_1+j}}{2}\biggr) \quad\mbox{and}
\nonumber
\\[-8pt]
\\[-8pt]
\nonumber
c&=&\sum_{j=1}^{K_2}F_X^{-1}
\biggl(\frac{1+\alpha
_{K_1+j}}{2}\biggr),
\end{eqnarray}
it holds
%
\begin{eqnarray}\label{eq2.8}\qquad
&&\mathbb{P}(\tilde{T}\leq t)
\nonumber
\\[-8pt]
\\[-8pt]
\nonumber
&&\qquad=\sum_{(j_1,\ldots,j_{K_2})\in\{0,1\}
^{K_2}}\prod
_{i=1}^{K_2}\alpha_{K_1+i}^{j_i}(1-
\alpha _{K_1+i})^{1-j_i}F_A\Biggl(t-c-\sum
_{i=1}^{K_2}j_ib_i\Biggr),
\end{eqnarray}
where $F_A(\cdot)$ is the CDF of $A$. Given the CDF, the expected
values of test statistic $\tilde{T}$ under null distributions can be
calculated as follows:
\begin{longlist}[1.]
\item[1.] For Fisher's method, it holds
\[
\mathbb{E}(\tilde{T})=2K_1-2\sum_{j=1}^{K_2}
\biggl[\alpha_{K_1+j}\log\biggl(\frac
{\alpha_{K_1+j}}{2}\biggr)+(1-
\alpha_{K_1+j})\log\biggl(\frac{1+\alpha
_{K_1+j}}{2}\biggr)\biggr],
\]
while the expectation of the original $T$ is $\mathbb{E}(T)=2K_1+2K_2=2K$.
\item[2.] For Stouffer's method, it holds
\[
\mathbb{E}(\tilde{T})=\sum_{j=1}^{K_2}
\biggl[\alpha_{K_1+j}\Phi^{-1}\biggl(\frac
{\alpha_{K_1+j}}{2}\biggr)+(1-
\alpha_{K_1+j})\Phi^{-1}\biggl(\frac{1+\alpha
_{K_1+j}}{2}\biggr)\biggr],
\]
while the expectation of the original $T$ is $\mathbb{E}(T)=0$.
\end{longlist}
\end{thmm}

 Note that there are $2^{K_2}$ terms summation in the
right-hand side of equation~(\ref{eq2.8}), which may cause severe computing
problem when $K_2$ is large. However, when some $\alpha_i$ are equal,
the formula can be simplified. Without loss of generality, assume there
are $r\geq1$ different $p$-value thresholds $\{\beta_1,\ldots,\beta_r\}$
such that
%
\begin{eqnarray}\label{eq2.9}
\sum_{j=1}^{K_2}\mathbbm{1}_{\{\alpha_{K_1+j}=\beta_1\}}=n_1,
\ldots,\sum_{j=1}^{K_2}\mathbbm{1}_{\{\alpha_{K_1+j}=\beta_r\}}&=&n_r
\quad\mbox {and}
\nonumber
\\[-8pt]
\\[-8pt]
\nonumber
 \sum_{l=1}^r
n_l&=&K_2,
\end{eqnarray}
then by defining $f(j;n_l,\beta_l):=\frac{n_l!}{j!(n_l-j)!}\beta
_l^j(1-\beta_l)^{n_l-j}$ for $j=0,\ldots,n_l$ and $l=1,\ldots,r$, the
formula can be simplified as
%
\begin{eqnarray}
&&\mathbb{P}(\tilde{T}\leq t)\nonumber\\
&&\qquad=\sum_{j_1=0}^{n_1}
\cdots\sum_{j_r=0}^{n_r}\prod
_{l=1}^rf(j_l;n_l,
\beta_l)\\
&&\hspace*{68pt}\qquad\quad{}\times F_A\Biggl(t-c-\sum_{l=1}^rj_l
\biggl(F_X^{-1}\biggl(\frac{\beta_l}{2}
\biggr)-F_X^{-1}\biggl(\frac{1+\beta_l}{2}\biggr)\biggr)
\Biggr).\nonumber
\end{eqnarray}
Therefore, the summation is reduced from $2^{K_2}$ terms to $\prod_{l=1}^r (n_l+1)$ terms.

 From the above theorem one concludes that $\tilde{T}$ is a
biased estimator of the original $T$. This motivates the following two
stochastic imputation methods.
%

\subsection{Single random imputation method}\label{sec2.3}
It is well known that the mean imputation method will underestimate the
variance of $\{p_{K_1+j}\}_{j=1}^{K_2}$ [\citet{Rubin}].
Furthermore, Theorem \ref{th1} indicates that the test statistic $\tilde{T}$
from the mean imputation method is a biased estimator of the original
$T$. To avoid this problem, one can replace the mean by randomly
simulating $q_i$ and $r_i$ from $\operatorname{Uniform}(0,\alpha_i)$ and $\operatorname
{Uniform}(\alpha_i,1)$, respectively.

 Recall that for $j=1,\ldots,K_2$, $B_j=F^{-1}_X(\tilde
{p}_{K_1+j})$. The next theorem (proof left to Appendix \ref{appB2}) states that $B_j\sim X$ holds under the null hypothesis, that is,
$B_j$ and $X$ follow the same distribution.

\begin{thmm}\label{th2}
For $j=1,2,\ldots,K_2$, it holds
%
\begin{equation}\label{eq2.11}
B_j\sim X.
\end{equation}
\end{thmm}

 The following corollary is a simple consequence of the above theorem.
\begin{corol*}
For the single random imputation method, the following facts hold for
$\tilde{T}$:
\begin{longlist}[1.]
\item[1.] For Fisher's method, it holds $B_j\sim\chi^2_2$ and
therefore $\tilde{T}\sim\chi^2_{2K}$.
\item[2.] For Stouffer method, it holds $B_j\sim\mathrm{N}(0,1)$ and
therefore $\tilde{T}\sim\mathrm{N}(0,K)$.
\end{longlist}
\end{corol*}

Therefore, in this case, $\tilde{T}$ is an unbiased estimator
of $T$ defined in equation~(\ref{eq2.1}).
\subsection{Multiple imputation method}\label{sec2.4}
Although the single random imputation method allows the use of standard
complete-data meta-analysis methods,\vadjust{\goodbreak} it cannot reflect the sampling
variability from one random sample. The multiple imputation method (MI)
overcomes this disadvantage [\citet{Rubin}]. In MI, each
missing value is imputed $D$ times. Therefore, $\{\tilde{T}^l\}
_{l=1}^D$ is a sequence of test statistics which are defined as
%
\begin{equation}\label{eq2.12}
\tilde{T}^l = \sum_{i=1}^K
F_X^{-1}\bigl(\tilde{p}^l_i
\bigr)=A+\sum_{j=1}^{K_2}B_j^l\qquad
\mbox{for } l=1,\ldots,D
\end{equation}
with
%
\begin{equation}\label{eq2.13}
q^l_i\sim\operatorname{Uniform}(0,\alpha_i)\quad
\mbox{and}\quad r^l_i\sim\operatorname {Uniform}(
\alpha_i,1).
\end{equation}

The test statistic is defined as $\overline{T}=\frac{1}{D}\sum_{l=1}^D\tilde{T}^l$, which satisfies
\begin{eqnarray*}
\overline{T}&=&A+\sum_{j=1}^{K_2}\Biggl[
\Biggl(\frac{1}{D}\sum_{l=1}^D
F^{-1}_X\bigl(q_{K_1+j}^l\bigr)\Biggr)
\cdot\mathbbm{1}_{\{p_{K_1+j}<\alpha_{K_1+j}\}
}\\
&&\hspace*{44pt}{}+\Biggl(\frac{1}{D}\sum
_{l=1}^D F^{-1}_X
\bigl(r_{K_1+j}^l\bigr)\Biggr)\cdot\mathbbm{1}_{\{
p_{K_1+j}\geq\alpha_{K_1+j}\}}
\Biggr]
\\
&=& A+\sum_{j=1}^{K_2}\Biggl[\Biggl(
\frac{1}{D}\sum_{l=1}^D
W_j^l\Biggr)\cdot\mathbbm {1}_{\{p_{K_1+j}<\alpha_{K_1+j}\}}+\Biggl(
\frac{1}{D}\sum_{l=1}^D
V_j^l\Biggr)\cdot \mathbbm{1}_{\{p_{K_1+j}\geq\alpha_{K_1+j}\}}\Biggr]
\\
&=& A+\sum_{j=1}^{K_2}\bigl[
\overline{W}_j\cdot\mathbbm{1}_{\{
p_{K_1+j}<\alpha_{K_1+j}\}}+\overline{V}_j
\cdot(1-\mathbbm{1}_{\{
p_{K_1+j}<\alpha_{K_1+j}\}})\bigr]=A+\sum_{j=1}^{K_2}Z_j.
\end{eqnarray*}

 Since $Z_j=\overline{W}_j$ with probability\vspace*{1pt} $\alpha_{K_1+j}$
and $Z_j=\overline{V}_j$ with probability $1-\alpha_{K_1+j}$, $Z_j$ is
a mixture distribution of $\overline{W}_j$ and $\overline{V}_j$ and,
therefore, $\overline{T}-A$ is a mixture distribution of $\{\overline
{W}_j, \overline{V}_j, j=1,\ldots,K_2\}$.

 Note that $W_j^l$ and $V_j^l$ are independent and identically
distributed (i.i.d.) for fixed~$j$. Denoting by $(\mu_{W_j}, \sigma
^2_{W_j}), (\mu_{V_j},\sigma^2_{V_j})$ the mean and variance of $W_j^l$
and $V_j^l$, respectively, then by the central limit theorem one
concludes that for large enough $D>0$ it holds
\begin{eqnarray*}
\overline{W}_j&=&\Biggl(\frac{1}{D}\sum
_{l=1}^D W_j^l\Biggr)\sim
\mathrm{N}\biggl(\mu_{W_j}, \frac{\sigma^2_{W_j}}{D}\biggr)\quad \mbox{and}\\
\overline{V}_j&=&\Biggl(\frac{1}{D}\sum
_{l=1}^D V_j^l\Biggr)\sim
\mathrm{N}\biggl(\mu_{V_j}, \frac{\sigma
^2_{V_j}}{D}\biggr).
\end{eqnarray*}

 Then the following theorem holds.

\begin{thmm}\label{th3}
For $(j_1,\ldots,j_{K_2})\in\{0,1\}^{K_2}$, by defining $U(j_1,\ldots,j_{K_2})=\sum_{i=1}^{K_2}(j_i\overline{W}_i+(1-j_i)\overline{V}_i)$
which satisfies
%
\begin{eqnarray}
&&U(j_1,\ldots,j_{K_2})
\nonumber
\\[-8pt]
\\[-8pt]
\nonumber
&&\qquad\sim \mathrm{N}\Biggl[\sum
_{i=1}^{K_2}\bigl(j_i\mu
_{W_i}+(1-j_i)\mu_{V_j}\bigr),\frac{1}{D}
\sum_{i=1}^{K_2}\bigl(j_i\sigma
^2_{W_i}+(1-j_i)\sigma^2_{V_j}
\bigr)\Biggr],
\end{eqnarray}
then for sufficiently large $D$, it holds approximately that
%
\begin{eqnarray}
\qquad&&\mathbb{P}(\overline{T}\leq t)
\nonumber
\\[-8pt]
\\[-8pt]
\nonumber
&&\qquad=\sum_{(j_1,\ldots,j_{K_2})\in\{0,1\}
^{K_2}}\prod
_{i=1}^{K_2}\alpha_i^{j_i}(1-
\alpha_i)^{1-j_i}\mathbb{P} \bigl(A+U(j_1,
\ldots,j_{K_2})\leq t\bigr).
\end{eqnarray}
The detailed notation is left to Appendix \ref{appC}.
\end{thmm}

 Similar to the mean imputation method, the formula can be
simplified when some $p$-value thresholds are equal, that is,
%
\begin{equation}\label{eq2.16}
\mathbb{P}(\overline{T}\leq t)=\sum_{j_1=0}^{n_1}
\cdots\sum_{j_r=0}^{n_r}\prod
_{l=1}^rf(j_l;n_l,
\beta_l)\mathbb{P}\bigl(A+U(j_1,\ldots,j_r)
\leq t\bigr),
\end{equation}
with
$U(j_1,\ldots,j_r)=\sum_{l=1}^r(j_lF^{-1}_X(q_l)+(n_l-j_l)F^{-1}_X(r_l)), q_l\sim\operatorname
{Uniform}(0,\beta_l)$ and $r_l\sim\operatorname{Uniform}(\beta_l,1)$.
\section{Simulation results}\label{sec3}
\subsection{Simulated expression profiles}\label{sec3.1}
To evaluate performance of the proposed imputation methods in the
genomic setting, we simulated expression profiles with correlated gene
structure and variable effect sizes as follows:

\textit{{Simulate gene correlation structure for $G=10\textup{,}000$ genes,
$N=100$ samples in each study and $K=10$ studies. In each study,
$4000$ of the $10\textup{,}000$ genes belong to $C=200$ independent clusters.}}
\begin{longlist}[\textit{Step} 1.]
\item[\textit{Step} 1.] Randomly sample gene cluster labels of
$10\mbox{,}000$ genes
($C_g\in\{0,1,\break 2, \ldots,C\}$ and $1\leq g\leq G$), such that $C=200$
clusters each containing $20$ genes are generated [$\sum_g \mathbbm
{1}(C_g=c)=20, \forall1\leq c\leq C=200$] and the remaining $6000$
genes are unclustered genes [$\sum_g\mathbbm{1}(C_g=0)=6000$].
\item[\textit{Step} 2.] For any cluster $c (1\leq c\leq C)$ in study $k (1\leq
k\leq K)$, sample $\Sigma'_{ck}\sim W^{-1}(\Psi,60)$,\vspace*{-1pt} where $\Psi
=0.5I_{20\times20}+0.5J_{20\times20}, W^{-1}$ denotes the inverse\break 
Wishart distribution, $I$ is the identity matrix and $J$ is the matrix
with all the entries being $1$. Set vector $\sigma_{ck}$ as the square
roots of the diagonal elements in $\Sigma'_{ck}$. Calculate $\Sigma
_{ck}$ such that $\sigma_{ck}\Sigma_{ck}\sigma_{ck}^T=\Sigma'_{gk}$.
\item[\textit{Step} 3.] Denote $g^{(c)}_{1},\ldots,g^{(c)}_{20}$ as the indices
for genes in cluster $c$. In other words, $C_{g^{(c)}_{j}}=c$, where
$1\leq c\leq200$ and $1\leq j\leq20$. Sample the expression of
clustered genes by $(X'_{g^{(c)}_{1}nk},\ldots,X'_{g^{(c)}_{20}nk})^T\sim \mathit{MVN}(0,\Sigma_{ck})$, where $1\leq n\leq
N=100$ and $1\leq k\leq K=10$. Sample the expression for unclustered
genes $X'_{gnk}\sim\mathrm{N}(0,1)$ for $1\leq n\leq N$ and $1\leq k\leq
K$ if $C_g=0$.
\end{longlist}
\textit{Simulate differential expression pattern.}
\begin{longlist}[\textit{Step} 4.]
\item[\textit{Step} 4.] Sample effect sizes $\mu_{gk}$ from $\operatorname{Unif}(0.1,0.5)
$ for $1\leq g\leq1000$ as DE genes and set $\mu_{gk}=0$ for
$1001\leq g\leq G$ as non-DE genes.
\item[\textit{Step} 5.] For the first $50$ control samples, $X_{gnk}=X'_{gnk}\ (1\leq g\leq G,1\leq n\leq N/2=50, 1\leq k\leq K)$. For cases,
$Y_{gnk}=X'_{g(n+50)k}+\mu_{gk}\ (1\leq g\leq G,1\leq n\leq N/2=50,
1\leq k\leq K)$.
\end{longlist}
 In the simulated data sets, $K=10$ studies with $G=10\mbox{,}000$
genes simulated. Within each study, there were $\frac{N}{2}=50$ cases
and $50$ controls. The first $1000$ genes were DE in all $10$ studies
with effect sizes randomly simulated from a uniform distribution on
$(0.1,0.5)$, respectively, and the remaining $9000$ were non-DE genes.
We chose this effect size range to produce an averaged standardized
effect size at $\frac{0.3}{1\cdot\sqrt{50}}= 0.1414$ so that the DE
analysis generates $\sim$500--600 candidate DE genes (Table~\ref{tab1}), a
commonly seen range in real applications. In each study, $200$ gene
clusters existed, each containing $20$ genes. The correlation structure
within each cluster was simulated from an inverse Wishart distribution.

 In the simulations, we performed a two sample $t$-test for each
gene in each study and then combined the $p$-values using the imputation
methods proposed in this paper. For simplicity, we viewed the $p$-values
from the last $5$ studies as truncated with thresholds 
$(\alpha_1,\ldots,\alpha_5)=(0.001,0.001,0.01,0.01,\break 0.05)$, respectively. In most genomic
meta-analysis, researchers often use conventional permutation analysis
by permuting sample labels to compute the $p$-values to preserve gene
correlation structure. However, such a nonparametric approach is not
applicable in our situation, since raw data are not available in some
studies. In order to control the false discovery rate (FDR), we
examined the Benjamini--Hochberg (B--H) method [\citet{Ben1}] and the Benjamini--Yekutieli (B--Y) method [\citet{Ben2}] separately. The number of DE genes detected at
nominal FDR rate $5\%$ were recorded and the true FDR rates were
computed for each meta-analysis method by
\[
\operatorname{FDR} = \frac{\sum_g \mathbbm{1}(\mathrm{gene\ \mathit{g}\ detected\ with\ }
g\geq1001)}{\# \{\mathrm{genes\ detected}\}}.
\]

 In the multiple imputation method, $D=50$ was selected.
Simulations were repeated for $50$ times and the mean and standard
errors of numbers of DE genes controlled by B--H and B--Y methods and their
true FDR are reported in Table~\ref{tab1}. The results showed that the FDRs were
controlled well for B--H correction but rather conservative for B--Y
correction (the true FDR of B--Y is only $1/10$ of B--H at nominal
$\operatorname{FDR}=5\%$). This is consistent with the previous observation that the
B--Y adjustment tends to be over-conservative since it guards against
any type of correlation structure [\citet{Ben2}]. As
a result, the B--H correction will be used for all applications
hereafter. The simulation results showed consistently that imputation
methods had higher statistical power than the available-case method,
and the mean imputation and multiple imputation methods outperform the
single random imputation method with similar performance. Surprisingly,
the ratio of detected DE genes compared to the complete case increased
from $41.6\%$ in the available case ($263.5/632.9$) to $80.4\%$ in mean
imputation ($508.6/632.9$) using Fisher's method. The improvement is
even more significant using Stouffer's method (from $41.8\%$ to $86.7\%$),
while at the same time the true FDRs were controlled at a similar
level for all methods. The result shows that imputation methods
successfully utilize the incomplete $p$-value information to greatly
recover the detection power.

\begin{table}
\tabcolsep=0pt
\caption{Simulation results for correlated data matrix at nominal
$\operatorname{FDR}=5\%$}\label{tab1}
\begin{tabular*}{\textwidth}{@{\extracolsep{4in minus 4in}}lcd{3.7}d{1.13}
d{3.7}c@{}}
\hline
& & \multicolumn{2}{c}{\textbf{Fisher}} & \multicolumn{2}{c@{}}{\textbf{Stouffer}}\\[-6pt]
& & \multicolumn{2}{c}{\hrulefill} & \multicolumn{2}{c@{}}{\hrulefill}\\
& \multicolumn{1}{c}{\textbf{Method/Mean (s.e.)}} &
\multicolumn{1}{c}{\textbf{No. DE}} & \multicolumn{1}{c}{\textbf{True FDR}} &
\multicolumn{1}{c}{\textbf{No. DE}} & \multicolumn{1}{c@{}}{\textbf{True FDR}} \\
\hline
B--H & Complete cases & 632.9\ (32.5) & 0.043\ (0.0013) & 518.6\ (36.2) &
0.046\ (0.0015)\phantom{00}\\
& Available-case & 263.5\ (37.4) & 0.048\ (0.0076) & 216.8\ (35.3) &
0.064\ (0.022)\phantom{000}\\
& Mean imputation & 508.6\ (35.1) & 0.046\ (0.0016) & 449.8\ (36.2) &
0.047\ (0.0022)\phantom{00}\\
& Single imputation & 408.9\ (35.7) & 0.043\ (0.0018) & 293.9\ (32.6) &
0.045\ (0.0027)\phantom{00}\\
& Multiple imputation & 509.2\ (35.0) & 0.045\ (0.0015) & 463.8\ (35.7) &
0.050\ (0.0019)\phantom{00}\\[3pt]
B--Y & Complete cases & 354.0\ (34.4) & 0.0041\ (0.00083) & 261.7\ (33.9) &
0.0036\ (0.00097)\\
& Available-case & 102.4\ (21.9) & 0.0047\ (0.0012) & 82.8\ (20.6) &
0.0029\ (0.00096)\\
& Mean imputation & 234.5\ (32.1) & 0.0037\ (0.00074) & 203.8\ (30.8) &
0.0034\ (0.00073)\\
& Single imputation & 164.0\ (27.3) & 0.0057\ (0.0014) & 113.5\ (22.3) &
0.0039\ (0.0015)\phantom{0}\\
& Multiple imputation & 235.3\ (32.0) & 0.0037\ (0.00075) & 216.1\ (30.9) &
0.0050\ (0.0010)\phantom{0}\\
\hline
\end{tabular*}
\end{table}

 We further examined the situation when gene dependence
structure does not exist [i.e., steps 1--3 were skipped and
$X'_{gnk}\sim\mathrm{N}(0,1)$]. Table~\ref{tab2} shows the true type I error
control under nominal significance level $5\%$ (i.e., true type I error
$= \frac{\sum_{g=1001}^{10{\scriptsize{\textup{,}}}000}\mathbbm{1}
(\mathrm{gene\ \mathit{g}\ is\ detected\ at\ significance\ level\ 0.05})}{9000}$). The result shows
adequate type I error control and confirms the validity of the closed
form or approximated formula of different imputation methods in Section~\ref{sec2}.

\begin{table}
\caption{Type I error control for independent data matrix at nominal
significance level $5\%$}\label{tab2}
\begin{tabular*}{\textwidth}{@{\extracolsep{\fill}}lcc@{}}
\hline
& \textbf{Fisher} & \multicolumn{1}{c@{}}{\textbf{Stouffer}} \\
\hline
Complete cases & 0.050 (0.00031) & 0.050 (0.00037)\\
Available-case & 0.050 (0.00035) & 0.050 (0.00033)\\
Mean imputation & 0.050 (0.00031) & 0.050 (0.00033)\\
Single imputation & 0.050 (0.00032) & 0.051 (0.00032)\\
Multiple imputation & 0.050 (0.00031) & 0.051 (0.00031)\\
\hline
\end{tabular*}
\end{table}

 To investigate the impact of $D$ on the performance of the
multiple imputation method, simulations were performed for $D\in\{
20,30,50,100,150,200,\break 250,300, 500\}$. The result is shown in
Appendix \ref{Appendix A}, Figure \hyperref[fig3]{3}, which demonstrates that the performance of the multiple
imputation method is quite robust for different number of imputation
$D$. We use $D=50$ throughout this paper.


\subsection{Simulation from complete real data sets}\label{sec3.2}
\begin{figure}

\includegraphics{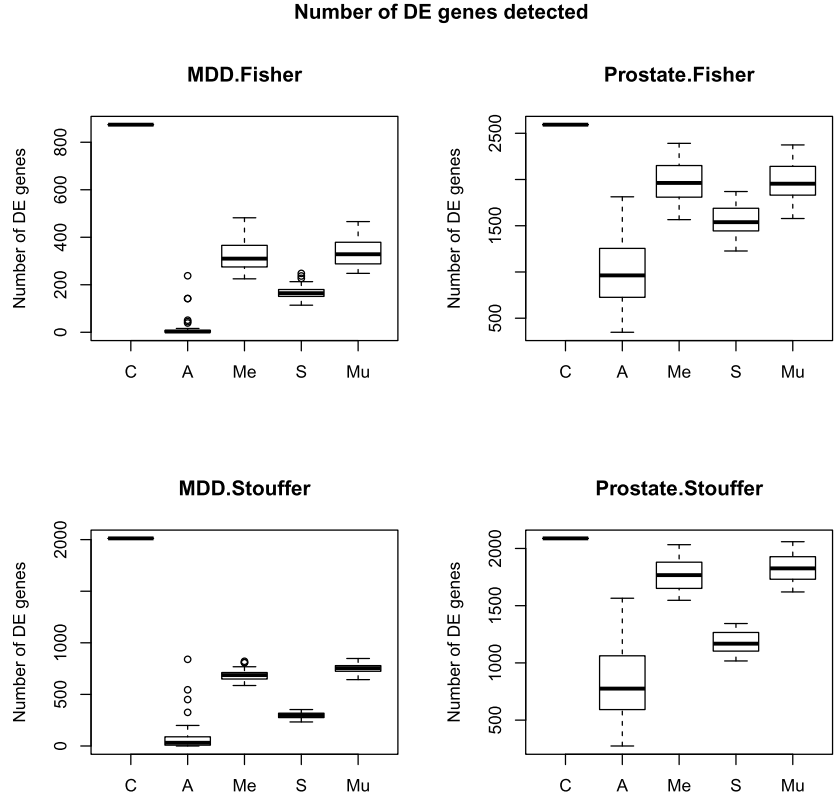}

\caption{Number of DE genes detected by Fisher's or Stouffer's method.
C: complete data; A: available-case; Me: mean-imputation; S:
single-imputation; Mu: multiple imputation.}\label{fig1}
\end{figure}

In this subsection the proposed methods were applied to two real
microarray data sets, including $7$ prostate cancer studies [Gorlov et al.
(\citeyear{Gorlov})] and $8$ major depressive disorder (MDD) studies
[\citet{Wang}]. The details are summarized in Supplement Table $1$
[\citet{supp}]. For each
data set, about half of the studies (four for MDD and three for
prostate cancer) were randomly selected with $p$-value truncation
threshold $0.05$. Five methods including complete data, available-case,
single random imputation, mean imputation and multiple imputation
methods were applied to the data sets with the simulated incomplete
data to impute by Stouffer's and Fisher's methods, respectively. The
generated $p$-values were corrected by the B--H method and the simulation
was repeated for $50$ times. Figure~\ref{fig1} shows boxplots of the numbers of
differentially expressed (DE) genes at $\operatorname{FDR}=1\%$ for different
methods in MDD and $\operatorname{FDR}=0.5\%$ for prostate cancer data.
Figure \ref{fig1} indicates similar conclusions that the multiple imputation and the
mean imputation methods detect more DE genes than the available-case
method and single random imputation method. In the MDD example, very
few DE genes (average of $16$ and $83$ for Fisher and Stouffer,
resp.) were detected using the available-case method if half of
the studies have truncated $p$-values. The mean and multiple imputation
methods greatly improved the detection sensitivity. About $95.2\%$
(Fisher) and $96.3\%$ (Stouffer) of DE genes detected by the mean
imputation method overlapped with DE genes detected by complete data
analysis in MDD and about $94.7\%$ (Fisher) and $88.1\%$ (Stouffer) of
DE genes detected by the mean imputation method overlapped with DE
genes detected by complete data analysis in prostate cancer, showing
the ability of imputation methods to recover DE gene detection power.\vadjust{\goodbreak}

\section{Applications}\label{sec4}

\subsection{Application to colorectal cancer}\label{sec4.1}
In the first motivating example, we followed Chan et al. (\citeyear{Chan}) and
attempted to collect 23 colorectal cancer versus normal gene expression
profiling studies. Raw data were available in only one study [Bianchini et al.
(\citeyear{Bianchini})] and $4$ of the other $22$ studies containing more than $100$ DE
genes at different $p$-value thresholds were included in our analysis. We
searched the GEO database and identified two additional new studies
[\citet{Jiang} and \citet{Bellot}]. The seven studies under
analysis were summarized in Table~\ref{tab3}. After gene-matching, $6361$ genes
overlapped in all three studies with raw data. The available-case\vadjust{\goodbreak}
method, the mean imputation method, the single random imputation method
and the multiple imputation method were applied for the seven studies
for the Fisher and Stouffer methods, respectively, and the results were
reported in Table~\ref{tab4}. For the single random imputation method and the
multiple imputation method, the analyses were repeated $50$ times and
the mean and standard error of the number of DE genes detected were
reported under FDR control by the B--H method. The results demonstrate
that for various FDR thresholds, the mean imputation method and the
multiple imputation method detected more DE genes than the
available-case method and the single random imputation method, which
was consistent with previous findings in simulations. Under $\operatorname
{FDR}=0.01\%$ control, the Fisher and Stouffer mean imputation detected
$2.07$ ($1183/571$) and $10.35$ ($383/37$) times of DE genes than those
by the available-case method, respectively.

\begin{table}
\tabcolsep=0pt
\caption{Seven colorectal cancer versus normal tissue expression
profiling studies included in analysis}\label{tab3}
\begin{tabular*}{\textwidth}{@{\extracolsep{\fill}}lcccccc@{}}
\hline
 & \textbf{No. of} &
\textbf{No. of} & \textbf{Raw data} & \textbf{No. of DE} & \textbf{No. of
overlapped} & $\bolds{p}$\textbf{-value}\\
\textbf{Study}& \textbf{samples}& \textbf{genes}& \textbf{availability} & \textbf{genes} & \textbf{DE genes} & \textbf{threshold}\\
\hline
Bianchini$\_2006$ & \phantom{0}24 & \phantom{0.}7403 & GSE3294 & -- & -- & --\\
Bellot$\_2012$ & \phantom{0}17 & 18,191 & GSE24993 & -- & -- & --\\
Jiang$\_2008$ & \phantom{0}48 & 18,197 & GSE10950 & -- & -- & -- \\
Grade$\_2007$ & 103 & 21,543 & -- & 1950 & 635 & 1e--7\\
Croner$\_2005$ & \phantom{0}33 & 22,283 & -- & \phantom{0}130 & \phantom{0}47 & 0.006\\
Kim$\_2004$ & \phantom{0}32 & 18,861 & -- & \phantom{0}448 & 143 & 0.001\\
Bertucci$\_2004$ & \phantom{0}50 & \phantom{0.}8074 & -- & \phantom{0}245 & \phantom{0}97 & 0.009\\
\hline
\end{tabular*}
\end{table}

\begin{table}[b]
\tabcolsep=0pt
\caption{Summary of results for colorectal cancer}\label{tab4}
\begin{tabular*}{\textwidth}{@{\extracolsep{\fill}}lcccccccc@{}}
\hline
& \multicolumn{4}{c}{\textbf{Fisher}} & \multicolumn{4}{c@{}}{\textbf{Stouffer}} \\[-6pt]
& \multicolumn{4}{c}{\hrulefill} & \multicolumn{4}{c@{}}{\hrulefill} \\
\textbf{FDR} & \textbf{Available} & \textbf{Mean} & \textbf{Single} &
\textbf{Multiple} & \textbf{Available} & \textbf{Mean} & \textbf{Single}
& \textbf{Multiple} \\
\hline
$1\%$ & 2587 & 2855 & 2172.4 (2.90) & 2785.4 (2.93) & 1318 & 1675 &
668.4 (3.96) & 1616.0 (2.10) \\
$0.1\%$ & 1472 & 1874 & 1265.6 (2.34) & 1805.7 (1.50) & \phantom{0}299 & \phantom{0}709 &
252.7 (1.93) & \phantom{0}680.5 (1.12)\\
$0.01\%$ & \phantom{0}571 & 1183 & \phantom{0}748.4 (1.89) & 1138.6 (2.00) & \phantom{00}37 & \phantom{0}383 &
102.5 (1.65) & \phantom{0}366.7 (0.69)\\
\hline
\end{tabular*}
\end{table}

\subsection{Application to pain research}\label{sec4.2}
The second motivating example comes from the meta-analysis of $20$
microarray studies of pain to detect the patterns of pain
[LaCroix-Fralish et al. (\citeyear{LaCroix})]. The original meta-analysis utilized DE gene
lists from each study under different threshold criteria from $p$-value,
FDR or fold change and identified $79$ ``statistically significant''\vadjust{\goodbreak}
genes that appeared in the DE gene lists of four or more studies. The
vote counting method essentially lost a tremendous amount of
information with flawed statistical inference. When we attempted to
repeat the meta-analysis, raw data of only one of the $20$ studies
(Barr$\_2005$) could be found. The old platform used in that study,
however, contained only $792$ genes and had to be excluded from further
meta-analysis. In the remaining $19$ studies, $11$ studies contained DE
gene lists under various $p$-value thresholds (marked bold in Supplement
Table~2 [\citet{supp}]) and were included in our application. In other words, this
example contained exclusively only studies with truncated $p$-values.
Table~\ref{tab5} shows the results of three imputation methods. Fisher and
Stouffer identified 280 and 45 genes under $5\%$ FDR control,
respectively. Note that the original meta-analysis tested the 79 genes
using an overall binomial test and the statistical significance was
controlled at an overall $p$-value level, not at a gene-specific FDR
level. As a result, DE gene lists from the new imputation methods are
theoretically more powerful and accurate.

 To validate the finding, we used the Gene Functional
Annotation tool from the DAVID Bioinformatics Resources website
(\url{http://david.abcc.ncifcrf.gov}). DAVID applied a modified Fisher's
exact test to evaluate the association between the DE gene lists and
pathways. Functional annotation of the 280 DE genes from the Fisher's
mean imputation method identified 208 pathways at $\operatorname{FDR}=5\%$, among
which selected important pain-related pathways were grouped into five
major biological categories and displayed in Table~\ref{tab6}. In contrast, the
79 genes from vote counting identified only 14 pathways, of which the
expected pain-related pathways under the categories of inflammation and
of differentiation, development and projection are missing (see Table~\ref{tab6}). The pathway enrichment $q$-values after multiple comparison control
of the ``280 gene list'' were very significant, while those of the ``79
gene list'' were not. Since the $p$-value calculation from Fisher's exact
test can be impacted by the DE gene size, we further compared the
enrichment odd-ratios of genes in the pathway versus in the DE gene
list. Still the enrichment odds-ratios of the ``280 gene list'' were
generally much higher than those for the ``79 gene list,'' showing
stronger pain functional association from the Fisher's mean imputation
method.\vadjust{\goodbreak}
%
\begin{table}
\caption{Summary of results for patterns of pain}\label{tab5}
\begin{tabular*}{200pt}{@{\extracolsep{\fill}}lcc@{}}
\hline
& \textbf{Fisher} & \textbf{Stouffer} \\
\hline
Mean & 280 & 45\\
Single & 57.04 (1.6228) & 16.44 (0.8605)\\
Multiple & 280.36 (0.8105) & 77.56 (0.6616)\\
\hline
\end{tabular*}
\end{table}
%

\begin{sidewaystable}
\tabcolsep=0pt
\caption{Summary of pathway analysis by DAVID}\label{tab6}
\begin{tabular*}{\textwidth}{@{\extracolsep{\fill}}llccd{2.1}cd{1.3}d{2.2}@{}}
\hline
& &  \multicolumn{3}{c}{\textbf{280 DE}} &
\multicolumn{3}{c@{}}{\textbf{79 DE}}\\
& & \multicolumn{3}{c}{\textbf{(Fisher's mean imputation)}} & \multicolumn
{3}{c@{}}{\textbf{(Vote counting)}} \\[-6pt]
& & \multicolumn{3}{c}{\hrulefill} & \multicolumn
{3}{c@{}}{\hrulefill} \\
 &  & &  & \multicolumn{1}{c}{\textbf{odds}} &  &  & \multicolumn{1}{c@{}}{\textbf{odds}} \\
\textbf{Category}& \multicolumn{1}{c}{\textbf{Pathway ID}}& \multicolumn{1}{c}{$\bolds{p}$\textbf{-value}}& \multicolumn{1}{c}{$\bolds{q}$\textbf{-value}}&
\multicolumn{1}{c}{\textbf{ratio}}& \multicolumn{1}{c}{$\bolds{p}$\textbf{-value}}&
\multicolumn{1}{c}{$\bolds{q}$\textbf{-value}}&
\multicolumn{1}{c@{}}{\textbf{ratio}} \\
\hline
Differentiation, & $\textbf{GO}:0030182\sim$ neuron differentiation &
5.6e--6 & 0.0006 & 3.1 & 0.26 & 0.95 & 1.6\\
development and & $\textbf{GO}:0045664\sim$ regulation of neuron
differentiation& 1.6e--5 & 0.0011 & 4.7 & 0.37 & 0.98 & 1.9\\
projection & $\textbf{GO}:0048666\sim$ neuron development & 2.5e--6 &
0.0003 & 3.6 & 0.24 & 0.94 & 1.7\\
& $\textbf{GO}:0051960\sim$ regulation of nervous system development&
6.5e--6 & 0.0006 & 4.2 & 0.29 & 0.96 & 1.9\\
& $\textbf{GO}:0031175\sim$ neuron projection development& 1.6e--5 &
0.0012 & 3.7 & 0.27 & 0.96 & 1.8\\
& $\textbf{GO}:0042995\sim$ cell projection & 3.6e--11 & 3.2e--9 & 3.5 &
0.033 & 0.47 & 1.9\\
& $\textbf{GO}:0043005\sim$ neuron projection & 3.0e--11 & 3.4e--9 & 4.3
& 0.043 & 0.51 & 2.0\\
& $\textbf{GO}:0030030\sim$ cell projection organization& 1.6e--5 &
0.0012 & 3.3 & 0.24 & 0.94 & 1.7\\[3pt]
Response  & $\textbf{GO}:0009611\sim$ response to wounding&
3.8e--10 & 2.8e--7 & 4.3 & 2.7e--5 & 0.016 & 3.6\\
to stimuli& $\textbf{GO}:0009719\sim$ response to endogenous stimulus& 3.2e--8 &
1.7e--5 & 3.4 & 0.35 & 0.97 & 1.3\\
& $\textbf{GO}:0048584\sim$ positive regulation of response to
stimulus& 7.9e--8 & 2.5e--5 & 4.9 & 0.0049 & 0.34 & 3.6\\
& $\textbf{GO}:0032101\sim$ regulation of response to external
stimulus& 1.1e--5 & 0.001 & 4.8 & 0.043 & 0.71 & 2.8\\[3pt]
Immune & $\textbf{GO}:0050778\sim$ positive regulation of immune
response& 4.2e--7 & 7.6e--5 & 5.9 & 0.018 & 0.57 & 4.0\\
& $\textbf{GO}:0002684\sim$ positive regulation of immune system
process& 1.9e--6 & 0.0003 & 4.4 & 0.0009 & 0.13 & 4.2\\
& $\textbf{GO}:0006956\sim$ complement activation & 3.0e--5 & 0.0016 &
11.5 & 0.011 & 0.46 & 8.4\\
& $\textbf{GO}:0002478\sim$ antigen processing and presentation  & 1.3e--6 & 0.00022 & 19.0 & 0.00098 & 0.12 &
10.64\\
&of
exogenous peptide antigen &&&&&&\\[3pt]
Inflammation & $\textbf{GO}:0002673\sim$ regulation of acute
inflammatory response & 1.4e--6 & 0.0002 & 14.1 & 0.19 & 0.93 & 3.8\\
& $\textbf{GO}:0002526\sim$ acute inflammatory response & 7.1e--6 &
0.0007 & 6.7 & 0.012 & 0.48 & 4.4\\
& $\textbf{GO}:0050727\sim$ regulation of inflammatory response &
1.9e--5 & 0.0012 & 6.9 & 0.17 & 0.92 & 2.8\\
& $\textbf{GO}:0006954\sim$ inflammatory response & 1.5e--5 & 0.0012 &
4.1 & 0.001 & 0.11 & 3.8\\[3pt]
Regulation of & $\textbf{GO}:0051969\sim$ regulation of transmission of
nerve impulse& 6.0e--6 & 0.0006 & 4.8 & 0.057 & 0.80 & 2.4\\
Transmission & & & & & & & \\
\hline
\end{tabular*}
\end{sidewaystable}
%

\subsection{Application to a three-way association method (liquid association)}\label{sec4.3}
So far the proposed imputation methods were applied successfully to two
real microarray data sets of colorectal cancer and pain research in
which the actual $p$-values of some genes were not reported in a subset
of studies. In this section we show that the proposed imputation
methods can be useful in the meta-analysis of ``big data'' such as GWAS
or eQTL, where the main computational problem is often the data
storage.

 In the literature it has long been argued that positively
correlated expression profiles are likely to encode functionally
related proteins. Liquid association (LA) analysis [Li (\citeyear{Li02})] is an
advanced three-way co-expression analysis beyond the traditional
pairwise correlations. For any triplet of genes $X, Y$ and $Z$, the LA
score $\mathit{LA}(X,Y|Z)$ measures the effect that expression of $Z$ to control
on and off of the co-expression between $X$ and $Y$. For example, high
expression of Z turns on positive correlation between $X$ and $Y$,
while when expression of $Z$ is low, $X$ and $Y$ are negatively or
noncorrelated. Theory in Li (\citeyear{Li02}) simplified calculation of the LA
score to a linear order of sample size and made the genome-wide
computation barely feasible. Supposing we want to combine $K$ studies
of the liquid association, liquid association $p$-values of all triplets
in all $K=10$ studies have to be stored for meta-analysis. When the
number of genes $G=1000$, the number of $p$-values to be stored is
$G\cdot C^{G-1}_2\cdot K=4.985$~GB. For a reasonable $G=20\mbox{,}000$
genome-wide analysis, storage size for all $p$-values quickly increases
to $39.994$~TB. One may argue that univariate (i.e., triplet by triplet)
meta-analysis may be applied repeatedly to avoid the need of storing
all $p$-value results. There are many other genomic meta-analysis
situations when this may not be feasible. For example, in GWAS
meta-analysis under a consortium collaboration, raw genotyping data
cannot be shared for privacy reasons and only the derived statistics or
$p$-values can be transferred for meta-analysis. Below we describe how
imputation methods can help circumvent the tremendous data storage
problem.

\begin{figure}

\includegraphics{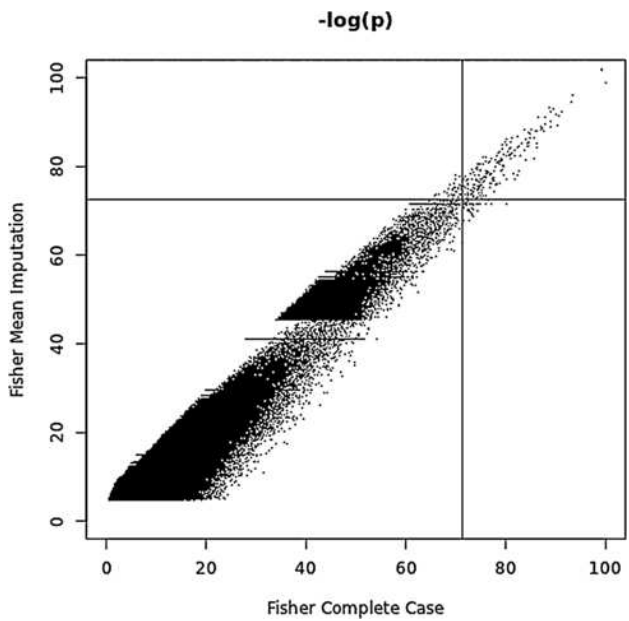}

\caption{$-\log(p)$ comparison of the mean imputation method using
truncated data with the complete case method using complete data.
Vertical line: $x=71.3$. Horizontal line: $y=72.58$. Points right to
vertical line are top $1000$ triplets detected by Fisher's complete
case method, and points above to horizontal line are top $1000$
triplets detected by Fisher's mean imputation method.}\label{fig2}
\end{figure}

 We performed a small scale of analysis on $566$ DE genes
previously reported from the meta-analysis of the eight MDD studies
used in Section~\ref{sec3.2} [\citet{Wang}]. The total number of possible
triplets $(X,Y|Z)$ was $90\mbox{,}180\mbox{,}780$. By setting up a $p$-value threshold
at $0.001$, we only needed to store exact $p$-values for
$2\mbox{,}094\mbox{,}123$ ($\sim$2.32\%) triplets and the remaining were truncated as considered
in this paper. Since we also needed to store the truncation index
information, we only needed to store $2\times2.32\%=4.64\%$ of the
information and the compression ratio was $ 95.36\%$. To investigate
the loss of information by the truncation, Figure~\ref{fig2} shows meta-analysis
$p$-values [at $-\log(p)$ scale] from Fisher's method using full data and
the Fisher mean imputation method using truncated data. The result
shows high concordance in the top significant triplets, which are the
major targets of this exploratory analysis. Among the top $1000$
triplets detected by Fisher's method using complete $p$-value
information, $83.7\%$ of them were also identified by the top $1000$ by
Fisher mean imputation. The remaining $163$ triplets were still in top
ranks (rank between $1199$ and $4763$) using truncated data in the
result of Fisher mean imputation. This result suggests a good potential
of applying data truncation to preserve the most informative
information and performing imputation to approximate the finding of the
top targets when meta-analysis of ``big data'' is needed. The compression
ratio may further increase by a more stringent truncation threshold,
but the performance may somewhat decline as a trade-off.

\section{Discussion and conclusion}\label{sec5}
When combining multiple genomic studies by $p$-value combination methods,
the raw data are often not available and only the ranges of $p$-values
are reported for some studies in genomic applications. This is
especially true for microarray meta-analysis since owners of many
microarray studies tend not to publish their data in the public domain.
This incomplete data issue is often encountered when one attempts to
perform a large-scale microarray meta-analysis. If raw data are not
available, two na\"{\i}ve methods---vote counting method and
available-case method---are commonly used. Since these two methods
completely or largely neglect the information contained in the
truncated $p$-values, their statistical power is generally low. In this
paper, we proposed three imputation methods for a general class of
evidence aggregation meta-analysis methods to combine independent
studies with truncated $p$-values: mean imputation, single random
imputation and multiple imputation methods. For each proposed
imputation method, the null distribution was derived analytically for
the Fisher and Stouffer methods. Theoretical results showed that the
test statistics from the single random imputation and the multiple
imputation methods were unbiased, while those for mean imputation
methods were biased. Simulations were performed for the imputed Fisher
method and imputed Stouffer method. The simulation results showed that
type I errors were well controlled for all methods, which was
consistent with our theoretical derivation. Compared to the naive
available-case method, all the imputation methods achieved higher
statistical powers, and the mean imputation and the multiple imputation
methods recovered much of the power that the complete cases method
achieved even when half of the studies had truncated $p$-values.
Furthermore, Figure \hyperref[fig3]{3} in Appendix \ref{Appendix A} showed that the power of the
multiple imputation method was robust to the number of imputation $D$.
Although small to moderate $D$ provided good results, we recommend
choosing $D$ being larger than $50$ to guarantee that the central limit
theorem can approximate well. Applications to two motivating examples
in colorectal cancer and pain conditions showed that both mean
imputation and multiple imputation performed among the best in terms of
detection sensitivity and biological validation by pathway analysis.

 In regression-type missing-data imputation methods, the null
distribution of the error term is unknown and is assumed to be normally
distributed with equal variance, a setting in which the multiple
imputation method usually outperforms the mean imputation in practice
and in theory [\citet{Rubin}], particularly because mean
imputation underestimates the true variance. However, our simulation
results demonstrated that the power of the two methods were quite
similar. Two reasons may contribute to this result. First, although the
test statistic from the mean imputation method is biased and neglects
the variation of truncated $p$-values, its $p$-value can be computed
accurately when the null distribution is derived analytically. Second
and more importantly, we find that the test statistic of mean
imputation is in fact $F^{-1}_X(\mathbb{E}(p))$, while for
sufficiently large $D$, the test statistic of multiple imputation
converges to $\mathbb{E}(F^{-1}_X(p))$ in distribution. It is easy to
show that these two quantities are very close to each other for a small
range of $p$, provided $F^{-1}_X(\cdot)$ is smooth. Since
$F^{-1}_X(\cdot)$ is infinitely differentiable for the Fisher and
Stouffer methods, and the small $p$-value range in $(0,\alpha)$ is
particularly of interest to us, it is not surprising that the mean
imputation method and multiple imputation method perform similarly.
Since the mean imputation method achieved almost the same power as the
multiple imputation method with less computational complexity, it is
more appealing and is recommended for microarray meta-analysis, where
the imputed meta-analysis method is performed repeatedly for thousands
of genes. In this paper only the evidence aggregation meta-analysis
methods are investigated and further work will be needed to extend
these results to order statistic based methods such as minP and maxP.\vadjust{\goodbreak}

 Note that although the truncated $p$-value issue discussed in
this paper may appear similar to the problem of ``publication bias,'' it
is fundamentally different. Publication bias refers to the fact that a
study with a large positive treatment effect is more likely to be
published than a study with a relatively small treatment effect,
resulting in bias if one only considers published studies. Denote by
$p_1,p_2,\ldots,p_N$ the $p$-values of all conducted studies that should
have been collected. Only a subset of likely more significant $p$-values
$p_1,p_2,\ldots,p_n $ are observed. Under this setting, $N$ is unknown
and $p_{n+1},\ldots,p_N$ are unknown as well. Since the number of
missing publications is unknown, Duval and Tweedie proposed the ``Trim
and Fill'' method to identify and correct for funnel plot asymmetry
arising from publication bias
[Duval and Tweedie (\citeyear{Susan1}) and (\citeyear{Susan2})], in
which an estimate of the number of missing studies is provided and an
adjusted treatment effect is estimated by performing a meta-analysis
including the imputed studies. For the truncated $p$-value problem we
consider here, the total number of studies, the number of studies with
truncated $p$-values and the $p$-value truncation thresholds are all known.
Therefore, investigation of the imputation of truncated $p$-values in
meta-analysis is different from the traditional ``publication bias''
problem and has not been studied in the meta-analysis literature, to
the best of our knowledge.

 In this paper the methods we developed mainly target on
microarray meta-analysis, but the issue can happen frequently in other
types of genomic meta-analysis [e.g., GWAS; Begurn et al. (\citeyear{Be02})]. In
Section~\ref{sec4.3} we demonstrated an unconventional application of our
methods to meta-analysis of liquid association. Due to the large number
of triplets tested in the three-way association, the needed $p$-value
storage is huge. By preserving only the most informative data by
truncation, the storage burden is greatly alleviated and our imputation
methods help approximate and recover the top meta-analysis targets with
little power loss. In an ongoing project, we also attempt to combine
multiple genome-wide eQTL results via meta-analysis. In eQTL,
regression analysis is used to investigate the association of a SNP
genotyping and a gene expression. It is impractical to store all
genome-wide eQTL $p$-values, as the storage space required is too large
($25\mbox{,}000$ genes $\times$ $2\mbox{,}000\mbox{,}000$ SNPS $= 5\times10^{10}$
$p$-values). A practical solution is to record only the eQTL $p$-values
smaller than a threshold (say, $10^{-4}$) for meta-analysis, which
leads to the same statistical setting as discussed in this paper. In
another project we combine results from multiple ChIP-seq peak calling
algorithms to develop a meta-caller. Since each peak caller algorithm
can only report the top peaks with\vadjust{\goodbreak} $p$-values smaller than a certain
$p$-value threshold, we again encounter the same truncated $p$-value
problem in meta-analysis. As more and more complex genomic data are
generated and the need for meta-analysis increases, we expect the
imputation methods we propose in this paper will find even more
applications in the future.\vspace*{1pt}

\begin{appendix}\label{app}
\section{Supplementary figure}\label{Appendix A}\vspace*{8pt}
%


\includegraphics{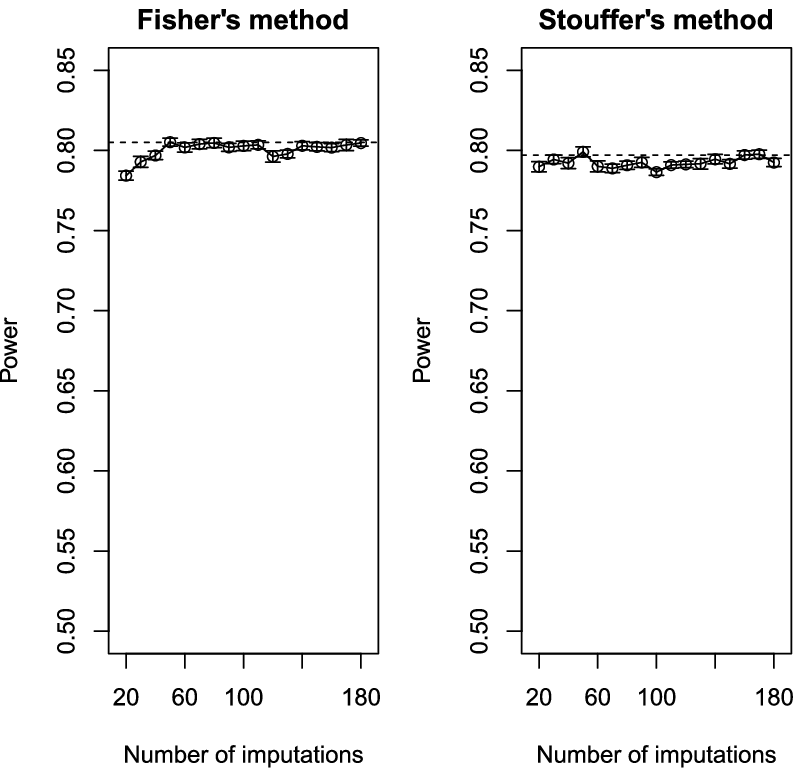}
\label{fig3}\vspace*{6pt}\\
 {\fontsize{9}{11}{\selectfont{\noindent\textsc{Fig. 3.}\hspace*{10pt}\textit{Power analysis at significance level $0.05$ for different
numbers\vspace*{-2pt} of imputation $D$. The dashed lines represent the theoretical asymptotic
power obtained by setting $D = 1000$.}}}}\vspace*{3pt}



\section{Proofs of theorems}\label{appB}
\subsection{Proof for Theorem~\texorpdfstring{\protect\ref{th1}}{1}}\label{appB1}
 Note that in this case, for $j=K_1+1,\ldots,K$, it holds
%
\begin{equation}
B_j=F_X^{-1}\biggl(\frac{\alpha_j}{2}\biggr)
\cdot\mathbbm{1}_{\{p_i<\alpha_j\}
}+F_X^{-1}\biggl(
\frac{1+\alpha_j}{2}\biggr)\cdot\mathbbm{1}_{\{p_i\geq\alpha_j\}}.
\end{equation}
Let $Y_j\sim\operatorname{Bernoulli}(\alpha_j)$. Since $p_i \sim\operatorname
{Uniform}(0,1)$ under the null hypothesis, it holds
%
\begin{eqnarray}
B_j&=&\biggl[F_X^{-1}\biggl(\frac{\alpha_j}{2}
\biggr)-F_X^{-1}\biggl(\frac{1+\alpha
_j}{2}\biggr)
\biggr]Y_j+F_X^{-1}\biggl(\frac{1+\alpha_j}{2}
\biggr)
\nonumber
\\[-8pt]
\\[-8pt]
\nonumber
&=&b_jY_j+c_j
\end{eqnarray}
and, therefore,
%
\begin{equation}
\tilde{T}=A+\sum_{j=1}^{K_2}b_jY_j
+c \qquad\mbox{with } c=\sum_{j=1}^{K_2}c_j.
\end{equation}

 For given $t$, it holds
%
\begin{eqnarray}
\mathbb{P}(\tilde{T}\leq t) &=& \mathbb{P}\Biggl(A+\sum
_{i=1}^{K_2}b_iY_i+c\leq t
\Biggr)\nonumber
\\
&=& \sum_{(j_1,\ldots,j_{K_2})\in\{0,1\}^{K_2}}\mathbb{P}\Biggl(A+\sum
_{i=1}^{K_2}b_iY_i+c\leq
t|Y_1=j_1,\ldots,Y_{K_2}=j_{K_2}\Biggr)\nonumber\\
&&\hspace*{70pt}{}\times\mathbb {P}(Y_1=j_1,\ldots,Y_{K_2}=j_{K_2})
\\
&=& \sum_{(j_1,\ldots,j_{K_2})\in\{0,1\}^{K_2}}\prod_{i=1}^{K_2}
\alpha _i^{j_i}(1-\alpha_i)^{1-j_i}
\mathbb{P}\Biggl(A\leq t-c-\sum_{i=1}^{K_2}j_ib_i
\Biggr)
\nonumber
\\
&=& \sum_{(j_1,\ldots,j_{K_2})\in\{0,1\}^{K_2}}\prod_{i=1}^{K_2}
\alpha _i^{j_i}(1-\alpha_i)^{1-j_i}F_A
\Biggl(t-c-\sum_{i=1}^{K_2}j_ib_i
\Biggr),
\nonumber
\end{eqnarray}
where $F_A(\cdot)$ is the CDF of $A$.

\subsection{Proof for Theorem~\texorpdfstring{\protect\ref{th2}}{2}}\label{appB2}

We show that for given $t$
%
\begin{eqnarray}
&&\mathbb{P}(B_i\leq t)\nonumber\\
 &&\qquad= \mathbb{P}\bigl(F^{-1}_X(
\tilde{p}_i)\leq t\bigr)= \mathbb{P}\bigl(\tilde{p}_i
\leq F_X(t)\bigr)
\nonumber
\\
&& \qquad=\mathbb{P}(x_i=1)\cdot\mathbb{P}\bigl(\tilde{p}_i
\leq F_X(t)|x_i=1\bigr)+\mathbb{P}(x_i=0)
\cdot\mathbb{P}\bigl(\tilde{p}_i\leq F_X(t)|x_i=0
\bigr)
\nonumber
\\[-8pt]
\\[-8pt]
\nonumber
&&\qquad= \alpha_i\mathbb{P}\bigl[q_i\leq
F_X(t)\bigr]+(1-\alpha_i)\mathbb{P}
\bigl[r_i\leq F_X(t)\bigr]
\\
&&\qquad= \cases{ %
\displaystyle\alpha_i\cdot
\frac{F_X(t)}{\alpha_i}=F_X(t), &\quad $\mbox{if } t\in \bigl(-\infty,F_X^{-1}(
\alpha_i)\bigr],$
\vspace*{2pt}\cr
\displaystyle\alpha_i+(1-\alpha_i)\cdot\frac{F_X(t)-\alpha_i}{1-\alpha_i}=F_X(t),
& \quad$\mbox{if } t\in\bigl(F_X^{-1}(\alpha_i),
\infty\bigr)$}
\nonumber
\\
&&\qquad= F_X(t),\nonumber
\end{eqnarray}
which implies that
%
\begin{equation}
B_i\sim X.
\end{equation}
%

\section{Some parameters in Theorem~\texorpdfstring{\lowercase{\protect\ref{th3}}}{3} for
The Stouffer and Fisher methods}\label{appC}

\subsection{Stouffer's method}

It is easy to obtain that
%
\begin{eqnarray}
\mu_{W_i}&=&\int_0^{\alpha}
\frac{1}{\alpha}\cdot\Phi^{-1}(t)\,dt=\frac
{1}{\alpha}\int
_{-\infty}^{\Phi^{-1}(\alpha)}u\,d\Phi(u)
\nonumber
\\[-8pt]
\\[-8pt]
\nonumber
&=&-\frac{1}{\alpha
\sqrt{2\pi}}e^{-{[\Phi^{-1}(\alpha)]^2}/{2}},
\\
\mu_{V_i}&=&\int_{\alpha}^1
\frac{1}{1-\alpha}\cdot\Phi^{-1}(t)\,dt=\frac
{1}{1-\alpha}\int
^{\infty}_{\Phi^{-1}(\alpha)}u\,d\Phi(u)\nonumber\\[-1pt]
&=&\frac
{1}{(1-\alpha_i)\sqrt{2\pi}}e^{-{[\Phi^{-1}(\alpha
)]^2}/{2}}
\nonumber\vspace*{-3pt}
\end{eqnarray}
and\vspace*{-2pt}
%
\begin{eqnarray}
\sigma^2_{W_i}&=& 1-\frac{\Phi^{-1}(\alpha)}{\alpha\sqrt{2\pi
}}e^{-{[\Phi^{-1}(\alpha)]^2}/{2}}-
\frac{1}{2\pi\alpha^2}e^{-[\Phi
^{-1}(\alpha)]^2},
\nonumber
\\[-9pt]
\\[-9pt]
\nonumber
\sigma^2_{V_i}&=& 1+\frac{\Phi^{-1}(\alpha)}{(1-\alpha)\sqrt{2\pi
}}e^{-{[\Phi^{-1}(\alpha)]^2}/{2}}-
\frac{1}{2\pi(1-\alpha
)^2}e^{-[\Phi^{-1}(\alpha)]^2}.\vspace*{-1pt}
\end{eqnarray}

\subsection{Fisher's method}
Similarly, it holds\vspace*{-1pt}
%
\begin{eqnarray}
\mu_{W_i} &=& \int_0^{\alpha}
\frac{1}{\alpha}\bigl(-2\ln(t)\bigr)\,dt=2[1-\ln\alpha ],
\nonumber
\\[-9pt]
\\[-9pt]
\nonumber
\mu_{V_i} &=& \int_{\alpha}^1
\frac{1}{1-\alpha}\bigl(-2\ln(t)\bigr)\,dt=2+\frac
{2\alpha}{1-\alpha}\ln(\alpha)\vspace*{-2pt}
\end{eqnarray}
and\vspace*{-2pt}
%
\begin{eqnarray}
\sigma^2_{W_i} &=& \mathbb{E}\bigl(W_i^2
\bigr)-\mu^2_{W_i}=4,
\nonumber
\\[-8pt]
\\[-8pt]
\nonumber
\sigma^2_{V_i} &=& \mathbb{E}\bigl(V_i^2
\bigr)-\mu^2_{V_i}=4-\frac{4\alpha
}{(1-\alpha)^2}\ln^2
\alpha.\vspace*{-2pt}
\end{eqnarray}
\end{appendix}

\section*{Acknowledgements}
The authors would like to thank S.~C. Morton for discussion. We wish to
express our sincere thanks to the Associate Editor and two reviewers
for their valuable comments to significantly improve this paper.\vspace*{-2pt}

\begin{supplement}[id=suppA]
\stitle{Supplementary tables}
\slink[doi]{10.1214/14-AOAS747SUPP} 
\sdatatype{.pdf}
\sfilename{aoas747\_Supp.pdf}
\sdescription{Supplementary Table 1: Detailed data sets descriptions;
Supplementary Table 2: $11$ pain-relevant microarray studies.}
\end{supplement}




\printaddresses
\end{document}